\documentclass[preprint,preprintnumbers,amsmath,amssymb, nofootinbib]{revtex4}

\usepackage{graphicx}
\usepackage{dcolumn}
\usepackage{bm}
\usepackage{epsfig}
\usepackage{color}
\usepackage{graphicx,subfigure}
\usepackage{enumerate}
\usepackage[bookmarks=true]{hyperref}

\usepackage{amsmath}
\usepackage{amssymb}

\def\fun#1#2{\lower3.6pt\vbox{\baselineskip0pt\lineskip.9pt
  \ialign{$\mathsurround=0pt#1\hfil##\hfil$\crcr#2\crcr\sim\crcr}}}

\input epsf

\newcommand{\be}{\begin{equation}}
\newcommand{\ee}{\end{equation}}
\newcommand{\bea}{\begin{eqnarray}}
\newcommand{\eea}{\end{eqnarray}}

\begin{document}
\global\long\def\met{\not{\!{\rm E}}_{T}}

\begin{flushright}
\vspace*{-1.5cm}ANL-HEP-PR-12-49\\
\vspace{-0.25cm} FERMILAB-PUB-12-323-T\\ 
\vspace{-0.25cm}NUHEP-TH/12-09
\end{flushright}

\vspace*{0.5cm}

\title{Have We  Observed the Higgs (Imposter)?}

\vspace*{0.2cm}

\author{
\vspace{0.5cm} 
Ian Low$^{a,b}$, Joseph Lykken$^{c}$, and Gabe Shaughnessy$^{d}$ }
\affiliation{
\vspace*{.2cm}
$^a$ \mbox{High Energy Physics Division, Argonne National Laboratory, Argonne, IL 60439}\\
$^b$ \mbox{Department of Physics and Astronomy, Northwestern University, Evanston, IL 60208} \\
$^c$  \mbox{Fermi National Accelerator Laboratory, P.O. Box 500, Batavia, IL 60510}\\
$^d$  \mbox{Department of Physics, University of Wisconsin, Madison, WI 53706}
\vspace*{0.8cm}}

\begin{abstract}
\vspace*{0.5cm}
We interpret the new particle at the Large Hadron Collider as a $CP$-even scalar and investigate its electroweak quantum number. Assuming an unbroken custodial invariance as suggested by precision electroweak measurements, only four possibilities are allowed if the scalar decays to pairs of gauge bosons, as exemplified by a dilaton/radion, a non-dilatonic electroweak singlet scalar, an electroweak doublet scalar, and electroweak triplet scalars. We show that current LHC data already strongly disfavor both the "plain-vanilla" dilatonic and non-dilatonic singlet imposters. On the other hand, a generic Higgs doublet give excellent fits to the measured event rates of the newly observed scalar resonance, while the Standard Model Higgs boson gives a slightly worse overall fit due to the lack signal in the $\tau\tau$ channel.  The triplet imposter exhibits some tension with the data.  The global fit indicates  the enhancement in the diphoton channel could be attributed to an enhanced partial decay width, while the production rates are consistent with the Standard Model expectations. We emphasize that more precise measurements of the ratio of event rates in the $WW$ over $ZZ$ channels, as well as the event rates in $b\bar{b}$ and $\tau\tau$ channels, are needed to further distinguish the Higgs doublet from the triplet imposter.

\end{abstract}


\maketitle

\section{Introduction}\label{sec:into}

The new resonance discovered \cite{CERNseminar} by the ATLAS and CMS experiments at the CERN Large Hadron Collider (LHC) could
be the long-sought Higgs boson of the Standard Model (SM) \cite{Weinberg:1967tq}. This is only the beginning of a challenging
 program of ``Higgs Identification'' to rigorously establish the quantum numbers and couplings of the new particle, and to reveal its
relationship, if any, to electroweak symmetry-breaking and fermion mass generation.

To confirm the identity of the new particle, we should first establish what it is not. For example, the diphoton decay mode shows
not only that the new state is a massive neutral boson but also that it does not have spin 1, which would violate the
Landau-Yang theorem \cite{Yang:1950rg}. By studying angular correlations in the decays to 4-lepton final states,
 it should be possible to distinguish whether the boson is $CP$ even, $CP$ odd, or a mixture \cite{Cao:2009ah, Gao:2010qx}, and eventually rule out
 the possibility that the boson has spin 2 rather than spin 0 \cite{Gao:2010qx}.
 
 Here we will assume that the new particle is a $CP$ even scalar, and address the question of determining its electroweak
 quantum numbers. A Higgs boson is the $CP$ even neutral component of a complex weak doublet with
 unit hypercharge, with the other three states comprising the Goldstone bosons that become the longitudinal
 components of the $W^{\pm}$ and $Z$ bosons. Together these four states also transform as a $({\bf 2}_L,{\bf 2}_R)$
 under the accidental $SU(2)_L \times SU(2)_R$ global symmetry of the SM lagrangian. 
After electroweak symmetry breaking, SM interactions still respect  an approximate diagonal symmetry called the custodial symmetry $SU(2)_C$ \cite{Sikivie:1980hm},
as evidenced by precision electroweak measurements of the $\rho$ parameter \cite{Nakamura:2010zzi}.

As shown in Ref.~\cite{Low:2010jp}, we can classify the leading order couplings of any neutral $CP$ even scalar
to $W$ and $Z$ bosons according to its properties under custodial symmetry. There are five possibilities that could
apply to the resonance discovered by ATLAS and CMS:

\begin{enumerate}[(1)]

\item The scalar is an electroweak singlet (and thus also a custodial singlet), but has dimension four couplings to $W$ and $Z$.
The latter property implies that the Higgs imposter is a dilaton \cite{Gildener:1976ih,Goldberger:2007zk}
or radion \cite{Goldberger:1999uk} resulting from
new electroweak symmetry-breaking dynamics in a strongly-interacting conformal sector or a warped extra dimension, the two being related by
AdS/CFT duality \cite{Maldacena:1997re}. The conformal dynamics couples the ``dilaton imposter'' $\chi$ to SM fermions, and to photons and gluons
through operators of dimension five.

\item The scalar is an electroweak singlet with dimension five couplings to $W$ and $Z$. This is the electroweak singlet imposter $s$
discussed in Ref.~\cite{Low:2011gn}. The dimension five couplings arise from integrating out other charged exotics,
which also generically produce dimension five couplings to photons and gluons, and higher dimension couplings
to SM fermions. This ``singlet imposter'' could be related to EWSB indirectly, either through an extended Higgs sector or as a "techni-axion" in technicolor models.

\item The scalar is not an electroweak singlet, but is  nevertheless a custodial singlet. This could be the Higgs boson $h$,
which is the custodial singlet component in the decomposition $({\bf 2}_L,{\bf 2}_R) = {\bf 1}\oplus{\bf 3}$. We will refer to this possibility simply as the ``Higgs boson,'' although it could very well be a custodial singlet in a more exotic representation of $SU(2)_L\times SU(2)_R$.

\item The scalar is the neutral member of a custodial 5-plet. This imposter could belong to an electroweak
triplet in an extended Higgs sector \cite{Georgi:1985nv}, and will be referred to as the ``triplet imposter'' $h_5$.

\item Mixtures of the above are possible. However note that, to the extent that mixtures (and thus mass eigenstates)
respect custodial symmetry, the only plausible possibility that cuts across cases is a mixture of (1) and (3) \cite{Giudice:2000av} or a mixture of (2) and (3) \cite{Barger:2007im}, as might indeed occur
in an extended Higgs sector. Mixtures that do not respect the custodial symmetry have been studied in Ref.~\cite{Logan:2010en}.

\end{enumerate}

There have been earlier works on fitting the couplings of a Higgs boson using the LHC 2011 data \cite{Carmi:2012yp}. In this work we wish to focus on understanding the electroweak property of the observed resonance. For simplicity we consider only the pure cases (1)-(4) for simplicity, and demonstrate that a "plain-vanilla" dilaton imposter in case (1), where all the SM gauge bosons including gluons and the electroweak gauge bosons are part of the conformal dynamics, as well as the singlet imposter in case (2) are already strongly disfavored by LHC data probing scalar couplings with pairs of SM gauge bosons $V_1V_2=\{WW, ZZ, Z\gamma, \gamma\gamma, gg\}$. We will show  that the custodial singlet Higgs in case (3) gives the best fit to current data and a SM Higgs boson, for which all couplings are fixed to the SM value, gives slightly worse fit. The triplet imposter in case (4) exhibits some tension with data, mainly due to the excess in $b\bar{b}$ and $\tau\tau$ channels.

\section{Scalar couplings to $V_1 V_2$}\label{sec:coupling}

As seen in Ref.~\cite{Low:2010jp},  tree level couplings to $W$ and $Z$  bosons of a scalar charged under electroweak symmetry  can be classified using the quantum number of the scalar under the custodial symmetry $SU(2)_C$, which is the diagonal subgroup,  after electroweak symmetry breaking,  of an accidental $SU(2)_L\times SU(2)_R$ global symmetry of the SM lagrangian.  The approximate 
custodial invariance  implies $\rho \equiv m_W^2/(m_Z^2 c_w^2) =1$, where $c_w$ is the cosine of the Weinberg angle, which was verified by the precision electroweak measurements to be true at  the precent level \cite{Nakamura:2010zzi}. 

The $SU(2)_L$ and the $U(1)_Y$ subgroup of $SU(2)_R$ is gauged in the SM, which implies that the weak isospin gauge bosons $W^a_\mu$ and the hypercharge gauge boson $B_\mu$ transform as a triplet and the $T^3$ component of a triplet, respectively, under $SU(2)_C$. Using the familiar rule for addition of angular momentum in quantum mechanics, it is immediately clear that a pair of $W/Z$ bosons can only couple to a $CP$ even neutral scalar that is either a custodial singlet $h$ or a custodial 5-plet $h_5$ (here both $h$ and $h_5$ are charged under $SU(2)_L\times U(1)_Y$).  Any $(\mathbf{N}_L, \mathbf{N}_R)$ representation of  $SU(2)_L\times SU(2)_R$ contains a custodial singlet  for $\mathbf{N}\ge 2$ and also a custodial 5-plet for $\mathbf{N}\ge 3$. The usual Higgs doublet scalar is a $(\mathbf{2}_L, \mathbf{2}_R)$ representation, while the $(\mathbf{3}_L, \mathbf{3}_L)=1\oplus 3 \oplus 5$ representation contains a real triplet scalar with $Y=2$ and a complex triplet scalar with $Y=0$.

We parameterize effective couplings  of  $h$ and  $h_5$ to $V_1V_2$ as:
\bea
\label{eq:hcoup}
{\cal L}_{hV_1V_2} &=&   c_V \left(\frac{2m_W^2}{v}\, h \,W^+_\mu W^{-\,\mu} +  \frac{m_Z^2}{v}\, h\, Z_\mu Z^{\mu}\right) \nonumber \\
&&\quad + c_g \frac{\alpha_s}{12\pi v} {h}\, G_{\mu\nu}^a G^{a\,\mu\nu} +c_\gamma \frac{\alpha}{8\pi v} {h}\, F_{\mu\nu} F^{\mu\nu} + c_{Z\gamma} \frac{\alpha}{8\pi v s_w}\,  {h}\, F_{\mu\nu} Z^{\mu\nu} \ ,\\
\label{eq:h5coup}
{\cal L}_{h_5V_1V_2} &=&   c_{5\,V} \left(-\frac{m_W^2}{v}\, h_5 \,W^+_\mu W^{-\,\mu} +  \frac{m_Z^2}{v}\, h_5\, Z_\mu Z^{\mu}\right) \nonumber \\
&&\quad + c_{5\,g} \frac{\alpha_s}{12\pi v} {h_5}\, G_{\mu\nu}^a G^{a\,\mu\nu} +c_{5\,\gamma} \frac{\alpha}{8\pi v} {h_5}\, F_{\mu\nu} F^{\mu\nu} + c_{5\,Z\gamma} \frac{\alpha}{8\pi v s_w} \,  {h_5}\, F_{\mu\nu} Z^{\mu\nu} \ ,
\eea
where $v\approx 246$ GeV.  The first lines in Eqs.~(\ref{eq:hcoup}) and (\ref{eq:h5coup}) contain  couplings to pairs of massive electroweak gauge bosons, which could arise at the tree level, while the second lines include couplings to massless gauge bosons (including the $Z\gamma$ channel), which only occur at one-loop level. Notice that ratios of couplings to $WW$ over $ZZ$ for the custodial singlet Higgs and the triplet imposter are different \cite{Low:2010jp}:
\be
\frac{g_{hWW}}{g_{hZZ}} = \frac{m_W^2}{m_Z^2} = c_w^2 \ , \qquad \frac{g_{h_5WW}}{g_{h_5ZZ}} = -\frac{m_W^2}{2m_Z^2} = -\frac{c_w^2}2 \ .
\ee
Otherwise they have similar coupling structure to $V_1V_2$. 

In the SM $c_g, c_\gamma$, and $c_{Z\gamma}$ are form factors which depend on the Higgs mass $m_h$, top quark mass $m_t$, and the $W$ boson mass $m_W$. More explicitly, 
\bea
c_g^{(SM)} &=& \frac34 A_{1/2}(\tau_t) \ , \\
c_\gamma^{(SM)} &=& A_1(\tau_W)+ N_c Q_t^2 A_{1/2}(\tau_t) \ , \\
c_{Z\gamma}^{(SM)} &=& 2\left[c_w A_1(\tau_W,\lambda_W) + N_c \frac{Q_t (2 T_3^{(t)} - 4 Q_t s_w^2)}{c_w} A_{1/2}(\tau_t,\lambda_t) \right]\ ,
\eea
where $N_c=3$ is the number of colors, $Q_t$ is the top quark electric charge in units of $|e|$,  $\tau_i=4m_i^2/m_h^2$, and $\lambda_i=4m_i^2/m_Z^2$. We use the same definitions of loop functions as in Ref.~\cite{Carena:2012xa}.  At 125 GeV, the numerical values are
\be
c_g^{(SM)}(125 \ {\rm GeV}) = 1 \ , \qquad c_\gamma^{(SM)}(125 \ {\rm GeV}) = -6.48 \ , \qquad c_{Z\gamma}^{(SM)}(125 \ {\rm GeV}) = -10.96 \ .
\ee
More generally, these coefficients would depend on the masses of new particles contributing to the decay widths. However, for on-shell production of the Higgs at a fixed mass, it is a good approximation to regard these coefficients as constant.

As mentioned in the Introduction, it is  possible to have a custodial singlet scalar that is also an electroweak singlet scalar, contrary to the Higgs boson $h$ which is charged under electroweak symmetry. For this possibility, the dilaton imposter $\chi$ turns out to have effective couplings to $V_1V_2$ that are identical to the ordinary Higgs boson \cite{Goldberger:2007zk}. So we have
\bea
\label{eq:chicoup}
{\cal L}_{\chi V_1V_2} &=&  c_{\chi\, V} \left(\frac{2m_W^2}{v}\, \chi \,W^+_\mu W^{-\,\mu} +  \frac{m_Z^2}{v}\, \chi\, Z_\mu Z^{\mu}\right) \nonumber \\
&&\quad + c_{\chi\,g} \frac{\alpha_s}{12\pi v} {\chi}\, G_{\mu\nu}^a G^{a\,\mu\nu} +c_{\chi\,\gamma} \frac{\alpha}{8\pi v} {\chi}\, F_{\mu\nu} F^{\mu\nu} + c_{\chi\,Z\gamma} \frac{\alpha}{8\pi v s_w}\,  {\chi}\, F_{\mu\nu} Z^{\mu\nu} \ .
\eea
Moreover, in the "plain-vanilla" scenario where all the SM gauge bosons are part of the conformal dynamics, the dilaton coupling to gauge bosons are determined entirely by the one-loop beta functions \cite{Goldberger:2007zk}, which would then predicts a dilaton-gluon-gluon coupling that is much enhanced over the SM values. We focus on the plain vanilla dialton in this work. In the other scenario, the singlet imposter $s$ discussed in case (2) in the Introduction, leading order couplings to all possible pairs of $V_1V_2$ come from dimension five operators and are induced only at the loop-level.  Three, and only three,  gauge-invariant operators could be generated at this order:
\be
\label{eq:Sope}
\kappa_{g} \frac{\alpha_s}{4\pi} \,\frac{s}{4 m_s} G_{\mu\nu}^a G^{a\, \mu\nu} +
\kappa_{W} \frac{\alpha}{4\pi s_w^2} \,\frac{s}{4 m_s} W_{\mu\nu}^a W^{a\, \mu\nu} +
\kappa_{B} \frac{\alpha}{4\pi c_w^2} \,\frac{s}{4 m_s} B_{\mu\nu} B^{\mu\nu} \ .
\ee 
At leading order these three operators determine the singlet coupling to all five pairs of SM gauge bosons, massive or not. In terms of mass eigenstates, the effective lagrangian for  couplings of a singlet imposter to SM gauge bosons is
\bea
\label{eq:singletcoup}
 {\cal L}_{sV_1V_2}& =&  \kappa_W \frac{\alpha}{8\pi m_s s_w^2}\, s \,W^+_{\mu\nu} W^{-\,\mu\nu} + \left(\kappa_W \frac{c_w^2}{s_w^2}+ \kappa_B \frac{s_w^2}{c_w^2}\right)  \frac{\alpha}{16\pi m_s}\, s\, Z_{\mu\nu} Z^{\mu\nu} \nonumber \\
&&+ \kappa_g \frac{\alpha_s}{16\pi m_s} {s}\, G_{\mu\nu}^a G^{a\,\mu\nu} + (\kappa_W+\kappa_B) \frac{\alpha}{16\pi m_s} {s}\, F_{\mu\nu} F^{\mu\nu} \nonumber \\
&& + \left(\kappa_W \frac{c_w}{s_w} - \kappa_B \frac{s_w}{c_w}\right) \frac{\alpha}{8\pi m_s}  {s}\, F_{\mu\nu} Z^{\mu\nu} \, ,
\eea
from which we see that, generically, couplings to the massive and massless gauge bosons are of the same order of magnitude, unlike other cases we considered so far where couplings to massive gauge bosons are tree level and the dominant decay channels. Expressions for the partial decay widths of the singlet scalar into SM gauge bosons can be found in Ref.~\cite{Low:2010jp}. From Eq.~(\ref{eq:singletcoup}) it is also clear that, if there is any change in the decay width in the diphoton channel, the partial width in the $Z\gamma$ channel would be modified as well \cite{Carena:2012xa}.\footnote{This statement is true generically, regardless of the electroweak quantum number of the scalar.}

As pointed out in Ref.~\cite{Low:2011gn} already, the democratic nature of a singlet imposter coupling to pairs of SM gauge bosons has important implications for phenomenology. First of all, the phase space factor now plays an important role in its decay patterns. For example, the phase space factor in the $gg$ channel is a factor of 8 larger than that in the diphoton channel because of color. Below kinematic thresholds decays into massive gauge bosons like $WW$ and $ZZ$ are suppressed generically, which is the case for the mass range we are interested in.  Moreover, decays into all four pairs of electroweak gauge bosons, $\{WW, ZZ, \gamma\gamma, Z\gamma\}$, are correlated with one another, as they are controlled by only two parameters, $\kappa_W$ and $\kappa_B$ from Eq.~(\ref{eq:singletcoup}).  In sharp contrast,  decays of $h$, $h_5$, or $\chi$ into $\gamma\gamma$ and $Z\gamma$ are controlled by two free parameters in Eqs.~(\ref{eq:hcoup}) and (\ref{eq:h5coup}), respectively, and  are independent of the decays into $WW$ and $ZZ$.

\section{Interpreting the data}\label{sec:data}

So far data collected at the LHC show the greatest sensitivities and significances in decay channels into $V_1V_2$, while there are also strong hints from decays into $b\bar{b}$ \cite{:2012cn} and, to a less extent, $\tau\tau$ final states. Before we present our analyses, it is worth recalling that what is being measured experimentally is the event rate $B\sigma_X(Y)$ for a particular production mechanism $X$ of the scalar $S=\{\chi, s, h, h_5\}$, which subsequently decays into final states $Y$:
\be
B\sigma_X(Y)\equiv \sigma(X\to S) \frac{\Gamma(S\to Y)}{\Gamma_{\rm tot}}\ ,
\ee
where $\Gamma_{\rm tot}$ is the total width of $S$. For  $V_1V_2$ channels at the LHC, two different production mechanisms are considered in current data: the gluon fusion $X=gg$ and the vector boson fusion (VBF) $X={\rm VBF}$, while three decay channels to gauge bosons are measured: $\{WW, ZZ, \gamma\gamma\}$. We will denote inclusive production of the scalar by  $X=pp$. The Tevatron $b\bar{b}$ result comes from the associated production of the Higgs with $W/Z$, $X=VH$. 
Experimental collaborations present their  $B\sigma_X(Y)$ in units of the SM  signal strength $B\sigma^{\rm (SM)}_X(Y)$ by defining a best-fit signal strength $\mu = B\sigma_X(Y)/B\sigma^{\rm (SM)}_X(Y)$. Given these notations, we consider the following results from the most recent LHC and Tevatron announcements as well as the 2011 LHC data:

\begin{enumerate}[(I)]

\item Inclusive channels 

\begin{enumerate}[(a)]

\item $B\sigma_{pp}(WW)$: 1.4$^{+0.5}_{-0.5}$  (ATLAS) \cite{ATLASWW}, 0.3$^{+1.1}_{-0.3}$ (Tevatron) \cite{:2012cn}.

\item $B\sigma_{pp}(ZZ)$: 1.1$^{+0.6}_{-0.4} $ (ATLAS 7 TeV) \cite{atlas2011},  0.7$^{+0.5}_{-0.4}$ (CMS)  \cite{cmshig12020}.

\item $B\sigma_{pp}(\gamma\gamma)$:  2.2$^{+0.7}_{-0.8} $ (ATLAS 7 TeV) \cite{atlas2011}, 1.8$^{+0.5}_{-0.8} $ (ATLAS 8 TeV) \cite{atlasgaga}, \\ 3.6$^{+3.0}_{-2.5}$ (Tevatron) \cite{:2012cn}.

\item $B\sigma_{pp}(\tau\tau)$: 0.5$^{+1.6}_{-2.1} $ (ATLAS 7 TeV) \cite{atlas2011}.

\end{enumerate}

\item Exclusive channels

\begin{enumerate}[(a)]

\item $B\sigma_{\rm Non-VBF}(\gamma\gamma)$: 1.7$^{+1.1}_{-1.1}$ (ATLAS) \cite{atlas2012}, 1.4$^{+0.6}_{-0.6}$ (CMS)  \cite{cmshig12020}.

\item $B\sigma_{\rm VBF}(\gamma\gamma)$:   2.8$^{+3}_{-2.3}$ \cite{atlas2012}, 2.2$^{+1.3}_{-1.1}$ (CMS)  \cite{cmshig12020}.
\item $B\sigma_{\rm Non-VBF}(WW)$:  0.7$^{+0.5}_{-0.5}$ (CMS)  \cite{cmshig12020}.
\item $B\sigma_{\rm VBF}(WW)$:  0.3$^{+1.5}_{-1.6}$ (CMS)  \cite{cmshig12020}.

\item $B\sigma_{\rm Non-VBF}(\tau\tau)$:  1.3$^{+1.1}_{-1.1}$ (CMS)  \cite{cmstautau}.
\item $B\sigma_{\rm VBF}(\tau\tau)$:  -1.8$^{+1.0}_{-1.0}$ (CMS)  \cite{cmstautau}.

\item $B\sigma_{VH}(b\bar{b})$: 0.5$^{+2.1}_{-2.2} $ (ATLAS 7 TeV) \cite{atlas2011},  0.5$^{+0.8}_{-0.8}$ (CMS)  \cite{cmshig12020}, \\2.0$^{+0.7}_{-0.7}$ (Tevatron) \cite{:2012cn}.

\end{enumerate}

\end{enumerate}
Unless otherwise stated, the LHC results assume combinations of the $\sqrt s=7$ and 8 TeV datasets.  ATLAS only provides results in inclusive channels, with the exception of $b\bar{b}$ channel. While CMS provides both inclusive and exclusive results, we only use the exclusive results in the fit so as to avoid double counting.

In the absence of  any information on the total width of the resonance, we could proceed in a model-independent fashion by  taking the ratios of event rates, so that the total width cancels in the ratio. On the other hand, if we make assumptions on the total width of the scalar, it is possible to fit the event rate itself, although the outcome is clearly model-dependent.

Taking ratios of event rates has the advantage that some of the common uncertainties, such as systematics and theoretical error in production cross section, should cancel \cite{Zeppenfeld:2000td}. In addition, modifications in  properties of the scalar that are universal in all decay channels would drop out in the ratio. Two examples are i) mixing with other scalars that have not been observed to date, and ii) higher dimensional operators giving additional contributions to the scalar kinetic term and resulting in a finite wave function renormalization of the scalar.\footnote{This is the effect of $c_H$ in the SILH lagrangian \cite{Giudice:2007fh}.} The drawback of taking the ratio, on the other hand, is that we may not have information on the overall normalization of the parameters in the effective lagrangian.

\subsection{Model-Independent Fits in $V_1V_2$ Channels}

We focus on taking ratio of event rates in diboson channels, since these provide useful discriminators among different Higgs imposters. Two classes of ratios could be taken:

\begin{itemize}

\item Ratios of event rates with the same production mechanism but different decay channels. In this class we consider:
\bea
{D}_{W/Z} &\equiv& \frac{B\sigma_{gg}(WW)}{B\sigma_{gg}(ZZ)} = \frac{\Gamma(S\to WW)}{\Gamma(S\to ZZ)} \ , \\
D_{\gamma/Z} &\equiv& \frac{B\sigma_{gg}(\gamma\gamma)}{B\sigma_{gg}(ZZ)} =  \frac{\Gamma(S\to \gamma\gamma)}{\Gamma(S\to ZZ)}\  , \\
D_{Z\gamma/Z} &\equiv& \frac{B\sigma_{gg}(Z\gamma)}{B\sigma_{gg}(ZZ)} =  \frac{\Gamma(S\to Z\gamma)}{\Gamma(S\to ZZ)} \ .
\eea
The first two ratios can be extracted from existing data, while the $Z\gamma$ decay channel has been suggested \cite{Gainer:2011aa}, but not reported. Since ATLAS did not report the exclusive channel, we use the number from the inclusive channel as an approximation. It is well-known that in the SM the inclusive rate is dominated by the $gg$ channel, with $VBF$ channel making up only about 7\% of the inclusive rate \cite{Djouadi:2005gi}; we include the relative weights of the $gg$ and $VBF$ production mechanisms when considering inclusive rates.

\item Ratios of event rates with different production mechanisms but the same decay channel. Since at the LHC the dominant production mechanisms are the $gg$ channel and, to a much lesser extent, the VBF channel, we only consider one ratio in this class:
\be
P_{g/V} \equiv \frac{B\sigma_{gg}(\gamma\gamma)}{B\sigma_{\rm VBF}(\gamma\gamma)}= \frac{\sigma(gg\to S)}{\sigma({\rm VBF}\to S)}\ .
\ee
When more data becomes available it will also be useful to form this ratio for the other three diboson channels.

\end{itemize}

\begin{figure}
\subfigure[]{
\includegraphics[scale=0.527, angle=0]{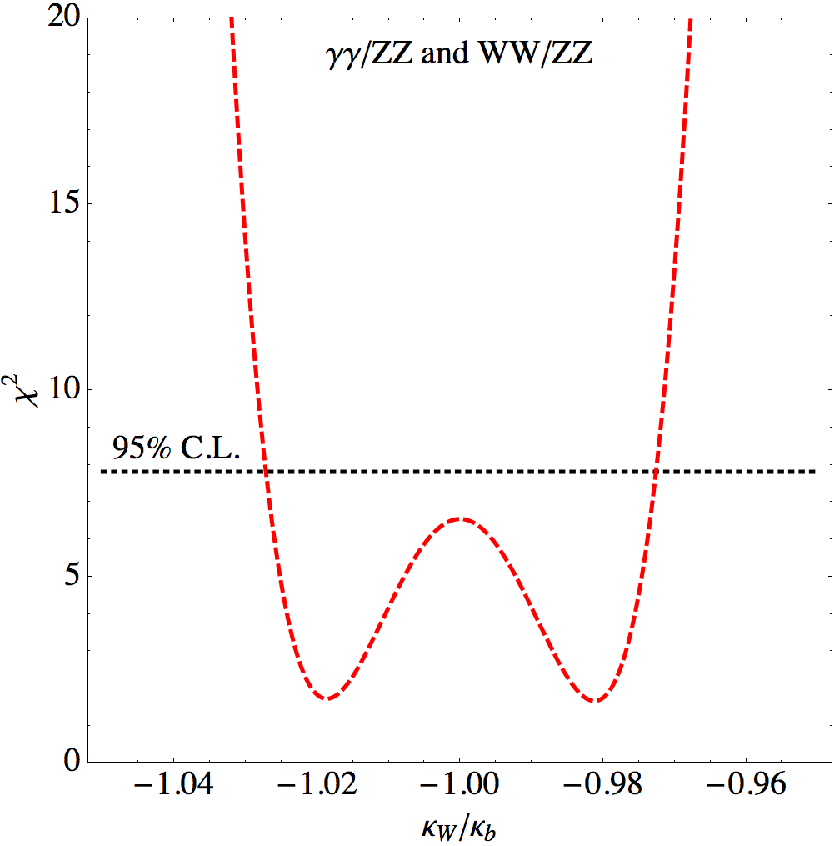}  \label{fig:1a}
}
\subfigure[]{
\includegraphics[scale=0.561, angle=0]{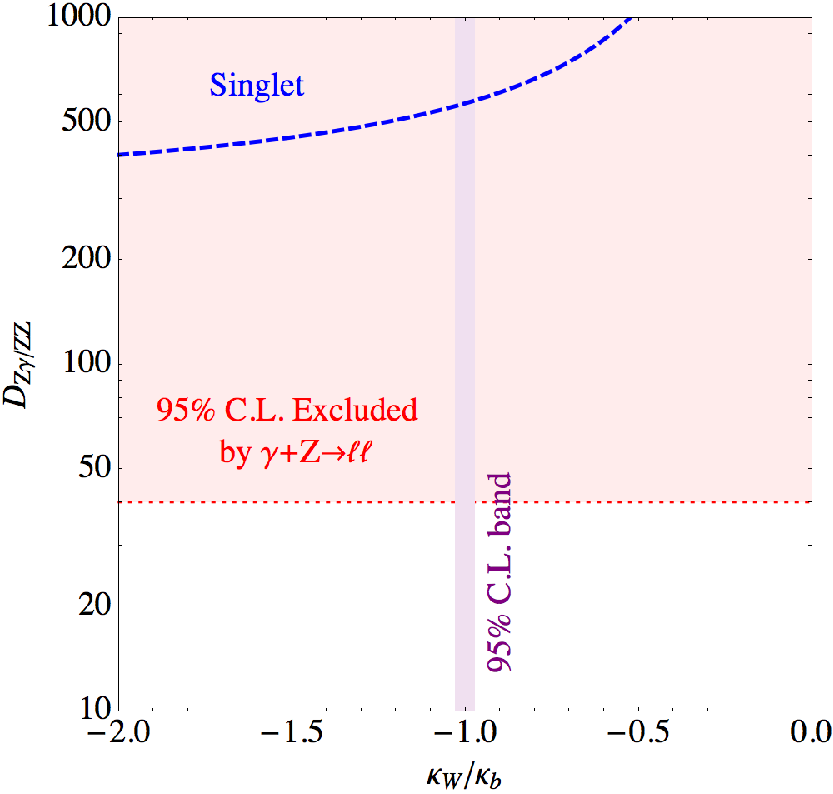}  \label{fig:1b}
            }
\caption{\label{fig:1}{\em  {\rm (a)} $\chi^2$ from fitting $D_{\gamma/Z}$ and $D_{W/Z}$ using one single parameter $\kappa_W/\kappa_B$, which is above the 95\% C.L. limit. {\rm (b)} The predicted $D_{Z\gamma/Z}$ using current data. The 95\% C.L. exclusion limit is derived from measurements of SM diboson production in the $Z\gamma$ channel, while the 95\% C.L. band for $\kappa_W/\kappa_B$ is derived from comparing $\Delta \chi^2$ with the best-fit value in {\rm (a)}.  }
}
\end{figure}

For a 125 GeV singlet imposter the decays into all four pairs of electroweak gauge bosons are controlled by only two free parameters, $\kappa_W$ and $\kappa_B$ in Eq.~(\ref{eq:singletcoup}). Therefore the  three ratios in the first class depend only on one number: $\kappa_W/\kappa_B$. In Fig.~\ref{fig:1a} we show  the $\chi^2$ of using one parameter $\kappa_W/\kappa_B$ to fit the measured $D_{\gamma/Z}$ and $D_{W/Z}$ from  ATLAS and CMS at the same time.  We see that the best-fit value is
\be
\label{eq:kwkbfit}
 \frac{\kappa_W}{\kappa_B} \approx -1 \ ,
\ee
and the absolute $\chi^2$ is below the 95\% C.L. limit. Using the above value, the predicted ratio of $D_{Z\gamma/Z}$ is
\be
D_{Z\gamma/Z} \sim 500 \ ,
\ee
which would be a spectacular signal. Although a dedicated search for a resonance in the $Z\gamma$ channel has not been reported, measurements for SM diboson production in the $Z\gamma$ channel have been made. Resonance decays in the $Z\gamma$ channel with a much enhanced rate certainly would contribute to this set of measurements as well. In Refs.~\cite{Aad:2011tc, Chatrchyan:2011rr} the event rates of  $\sigma(pp\to Z\gamma +X) \times Br(Z\to \ell^+\ell^-)$  are measured to be consistent with that expected from the SM prediction:
\bea
\text{ATLAS}&:& 6.5 \pm 1.2 {\rm (stat)} \pm 1.7 {\rm (syst.)} \pm 0.2 {\rm (lumi)}\  \text{pb}\ ,\qquad \text{Theory}: 6.9\pm 0.5 {\rm \ pb} \ , \nonumber \\
\text{CMS}&:& 9.4 \pm 1.0 {\rm (stat)} \pm 0.6 {\rm (syst.)} \pm 0.4 {\rm (lumi)}\  \text{pb}\ ,\qquad \text{Theory}: 9.6\pm 0.4 {\rm \ pb} \ . \nonumber 
\eea
The different values for ATLAS and CMS result from different selection cuts. On the other hand, using the best fit signal strength for $B\sigma_{pp}(ZZ)$ at the LHC, we see that the predicted $\sigma(pp\to s\to Z\gamma+X)\times Br(Z\to \ell^+\ell^-)\sim 15$ pb. Although we have not simulated the selection efficiency of the resonance decays into $Z\gamma$ for the cuts imposed in Refs.~\cite{Aad:2011tc, Chatrchyan:2011rr}, it is worth noting that the $p_T$ distribution of the photon from resonance decays is peaked at $m_s^2-m_Z^2/(2m_s) \approx 30$ GeV, while that from the SM diboson production is peaked at $p_T=0$ \cite{Gainer:2011aa}. Therefore, we expect a significant amount of the events from the resonance decay to pass the photon $p_T$ cut. In the end, we see that an event rate of the order of 15 pb in the resonance decays into $Z+\gamma\to \ell^+\ell^-+\gamma$ is strongly disfavored.  Using these arguments we  derive an estimate of the 95\% C.L. limit on $D_{Z\gamma/Z}$, using the measured $B\sigma_{pp}(ZZ)$, which is shown in Fig.~\ref{fig:1b}. We see that the predicted $D_{Z\gamma/Z}$ from a singlet imposter is an order of magnitude larger than the 95\% C.L. limit. Therefore, Fig.~\ref{fig:1} shows that a singlet imposter is excluded at 95\% C.L. as the interpretation of the excess at the LHC.

It is possible to understand why the partial width in the $Z\gamma$ channel is enhanced by so much for the  singlet imposter. As mentioned in the end of Sect.~\ref{sec:coupling}, its  couplings to gauge bosons are democratic and the partial width is largely determined by phase factors and kinematics. Therefore at 125 GeV, partial widths of $s$ has the following generic feature \cite{Low:2011gn}:
\be
\label{eq:pattern}
\Gamma_{gg}\  \agt\  \Gamma_{\gamma\gamma} \ \agt\  \Gamma_{Z\gamma}\  \agt \ \Gamma_{WW} \ \agt\  \Gamma_{ZZ}  \ .
\ee
However, in the SM we have
\be
\Gamma_{WW}^{(\rm SM)} \ >\  \Gamma_{gg}^{(\rm SM)}\  >\  \Gamma_{ZZ}^{(\rm SM)}\  >\  \Gamma_{\gamma\gamma}^{(\rm SM)}\  >\  \Gamma_{Z\gamma}^{(\rm SM)}\ ,
\ee
and current measurements  suggest a diphoton partial width that is still smaller than those in the $WW$ and $ZZ$ channels. Therefore the diphoton decay width of a singlet imposter should be suppressed from its generic expectation in order to fit the measured event rate. In Eq.~(\ref{eq:singletcoup}) the $s$-$\gamma$-$\gamma$ coupling is controlled by $\kappa_W+\kappa_B$, which explains why the best fit value is $\kappa_W/\kappa_B \approx -1$. In this region we see from Eq.~(\ref{eq:singletcoup}) that there is also a partial cancellation in the $s$-$Z$-$Z$ coupling, while  the $s$-$Z$-$\gamma$ coupling is enhanced. Together with the fact that at 125 GeV the $ZZ$ final state is below kinetic threshold, it is not surprising that the predicted $Z\gamma$ partial width is much larger than the $ZZ$ partial width.

\begin{figure}
\subfigure[]{
\includegraphics[scale=0.545, angle=0]{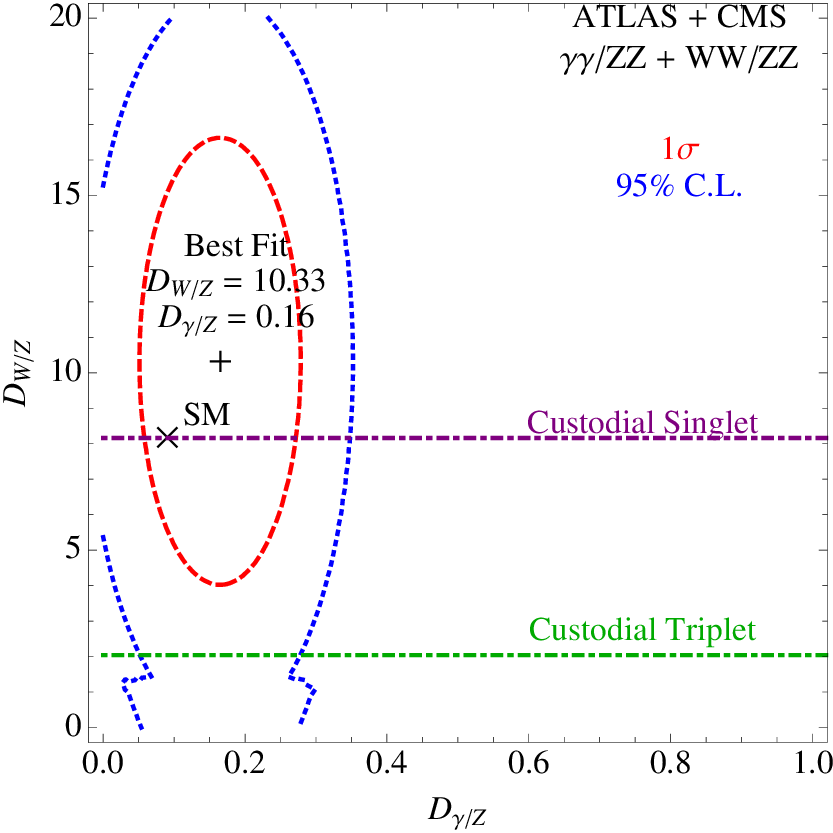}  \label{fig:2a}
}
\subfigure[]{
\includegraphics[scale=0.572, angle=0]{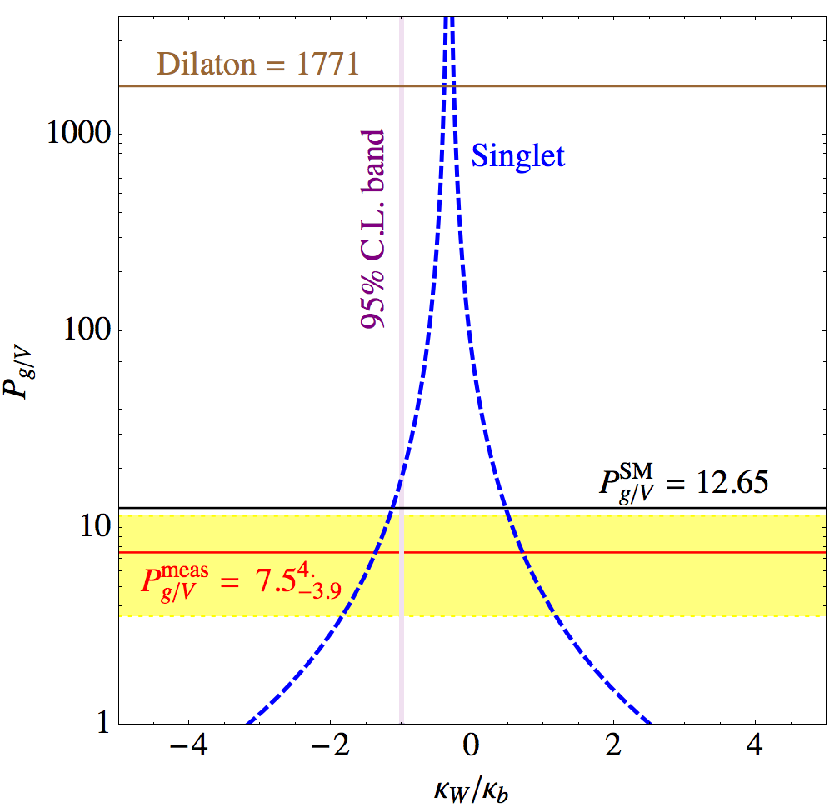}  \label{fig:2b}
            }
\caption{\label{fig:2}{\em  {\rm (a)}  Predictions of Higgs boson and the triplet imposter on the $D_{W/Z}$-$D_{\gamma/Z}$ plane. The custodial singlet is within $1\sigma$ range while the custodial triplet is within the $2\sigma$ range of the measured value.  The feature at the bottom of the $2\sigma$ contour is due to asymmetric uncertainties.
{\rm (b)} The predicted $P_{g/V}$ for the dilaton and the singlet imposters. The dilaton imposter predicts a $P_{g/V}$ that is strongly disfavored.}
}
\end{figure}

For a 125 GeV custodial singlet and 5-plet, $D_{W/Z}$ is completely fixed to be, 
\be
D_{W/Z}^{(h)} = 8.16\ , \qquad D_{W/Z}^{ (h_5)} = \frac14 D_{W/Z}^{(h)}=2.04\ ,
\ee
Thus a large deviation of  $D_{W/Z}$ from these two values would disfavor the custodial singlet Higgs and the triplet imposter as the interpretation of the excess.\footnote{One could  include higher dimensional operators which break custodial invariance to shift $D_{W/Z}$ away from the SM value for a Higgs boson \cite{Farina:2012ea}. However, a potentially large effect is needed,  implying a low cut-off for the higher dimensional operators and new light degrees of freedom at the electroweak scale, which may be in tension with null results from direct searches.} The ratio $D_{\gamma/Z}$ also allows for an estimate of the ratios $c_\gamma/c_V$ and $c_{5\,\gamma}/c_{5\,V}$. There is no prediction on the $D_{Z\gamma/Z}$ in these two scenarios,  although simultaneous measurements of $D_{\gamma/Z}$ and $D_{Z\gamma/Z}$ may shed light on  electroweak properties of new light degrees of freedom mediating scalar decays in the $\gamma\gamma$ and $Z\gamma$ channel \cite{Carena:2012xa}. In Fig.~\ref{fig:2a} we show the ratios extracted from the LHC data  on the $D_{W/Z}$--$D_{\gamma/Z}$ plane, as well as the expectations for the custodial singlet Higgs and the triplet imposter. We see that the custodial singlet Higgs is consistent with data within $1\sigma$ contour, while the triplet imposter is consistent  within the 95\% C.L. limit.

It should be emphasized that $\{WW, ZZ, \gamma\gamma\}$ are the three channels with the most sensitivity to Higgs boson searches. So the uncertainties in Fig.~\ref{fig:2a} could be reduced significantly in the future, which would then allow for better discrimination between the Higgs boson and the triplet imposter.

In Fig.~\ref{fig:2b} we show the extracted $P_{g/V}$ from data using only diphoton final states, which have the best precision, as well as the predicted ratio for the dilaton and the singlet imposters.  The SM expectations at 7 and 8 TeV are \cite{lhc-x-section}
\be
\label{eq:pgvsm}
P_{g/V}^{({\rm SM}) }({\rm 7 \ TeV})= 12.65 \ , \qquad P_{g/V}^{({\rm SM})} ({\rm 8 \ TeV})= 12.52\ .
\ee
However, while the ATLAS presented best-fit signal strengths in gluon fusion and VBF production channels in Ref.~\cite{atlas2012}, CMS only presented best-fit signal strengths in "VBF-tag" category, which is expected to have some $gg\to h$ contamination with the additional two jets arising from higher-order QCD effects.  To account for this contamination, we include a 25\% contribution from gluon fusion for the CMS VBF-tag signal strength:
\be
 \sigma(pp\to S+2j) = \epsilon\ \sigma(gg\to S+2j) + (1-\epsilon)\ \sigma({\rm VBF} \to S+2j) ,
\ee
where CMS assumes $\epsilon\sim 0.25$ \cite{cmsgaga}.  We can then relate CMS measured value of $P^{\rm VBF-tag}_{g/V}$ to the true $P_{g/V}$:
\be
P_{g/V}^{\rm VBF-tag} = \frac{P_{g/V}}{1+\epsilon \left(P_{g/V} -1\right)}\ ,
\ee
which can be used to obtained the true $P_{g/V}$ from the CMS measurements. Such a conversion is not needed for the ATLAS results since the numbers are presented in terms of production channels, not selection categories. In the end we find
\be
P_{g/V}^{\rm meas} = 7.5^{+4.0}_{-3.9}\ ,
\ee
which is the combined value for the ATLAS and CMS diphoton measurements. We see that the SM value in Eq.~(\ref{eq:pgvsm}) is consistent with the measured value. 
The  observation that
\be
 P_{g/V} \sim P_{g/V}^{({\rm SM}) }
\ee
suggests the enhancements in $B\sigma_{gg}(\gamma\gamma)$ and $B\sigma_{\rm VBF}(\gamma\gamma)$ could be explained simultaneously with an enhanced diphoton partial width resulting from an increased $c_\gamma$ or $c_{5\, \gamma}$. We will see that this is indeed the case when fitting the event rates directly.

It turns out that $P_{g/V}$ could be used as a discriminator for the "plain-vanilla" dilaton imposter \cite{Coleppa:2011zx}, which at 125 GeV gives
\be
P_{g/V}^{\rm (D)} = 140 \times P_{g/V}^{({\rm SM}) }  \sim 1700 \ ,
\ee
which is clearly disfavored strongly by current measurements. This prediction of ratio holds for the radion in Randall-Sundrum model as well \cite{Barger:2011hu}. Essentially a dilaton imposter is ruled out as soon as one can establish the presence of the VBF production channel. In Fig.~\ref{fig:2b} we show the ratio $P_{g/V}$ for the dilaton and singlet imposters, as well as the SM expectation and the value extracted from current data.


\subsection{Model-Dependent Fits in All Channels}

Since many significant cross section measurements have been made by the LHC and the Tevatron, we can fit the parameters  of model.  Since the dilaton and the singlet imposters can not fit the model-independent ratios considered in the previous subsection, we only consider the Higgs boson and the triplet imposter when fitting all channels. In order to include data in the $b\bar{b}$ and $\tau\tau$ channels, we need to introduce the Higgs couplings to $b\bar{b}$ and $\tau\tau$:
\be
{\cal L}_{hff} = c_b \frac{m_b}{v} h\bar{b}b +  c_\tau \frac{m_\tau}{v} h\bar{\tau}\tau\ ,
\ee
where $c_b^{\rm (SM)} = c_\tau^{\rm (SM)} = 1$. On the other hand, the triplet imposter does not have renormalizable couplings to SM fermions, so we simply set 
\be
{\cal L}_{h_5ff} = 0\ .
\ee
For the total width, we parametrize it as
\bea
\Gamma^{h}_{\rm tot} &=&  \sum_{V_1 V_2} \Gamma(h\to V_1 V_2) + \sum_{f} \Gamma(h\to f\bar f) \ , \\ 
\Gamma^{h_5}_{\rm tot} &=&  \sum_{V_1 V_2} \Gamma(h_5\to V_1 V_2)  \ . \
\eea
Therefore the total width depends on  all the $c$ coefficients in the effective couplings during the fit. In principle one could introduce an extra free parameter in the total width to incorporate the possibility that the scalar could decay into other channels that have not been observed. In the end, we fit five parameters, $\{c_g, c_V, c_\gamma, c_b, c_\tau\}$, for the Higgs boson and three parameters, $\{c_{5\, g}, c_{5\, V}, c_{5\, \gamma}\}$, for the triplet imposter.

\begin{table}[t]
\caption{\em Comparison of fits for a SM Higgs, a generic Higgs boson, and a triplet imposter.
One dimensional parameter estimates in the custodial singlet ($h_1$) and triplet ($h_5$) models under the total width assumptions.  Uncertainties indicate the $1\sigma$ range.  The SM Higgs boson is encapsulated in the custodial singlet scenario with $\Gamma_{\rm tot} = \Gamma_{\rm tot}^{h_{SM}}$, with $c_g=c_V=1$ and $c_\gamma=6.48$.   }
\begin{center}
\begin{tabular}{|c|cc|ccccc|}
\hline
   & $\chi^2/\nu$& $p$-value & $c_g $ & $c_V$ & $c_\gamma  $ & $c_b  $ & $c_\tau $ \\
\hline
SM Higgs & 1.08  & 0.63 & 1 & 1 & 6.48 & 1 & 1 \\
Higgs Boson & 0.74 & 0.27   &$0.92^{+0.30}_{-0.19}$ & $1.07^{+0.15}_{-0.17}$ & $9.7^{+1.9}_{-1.8}$ &  $1.1^{+0.5}_{-0.4}$ &  $<0.73$\\
Triplet Imposter & 1.34 & 0.84 & $0.37^{+0.08}_{-0.06}$ & $0.45^{+0.10}_{-0.09}$ & $3.8^{+0.5}_{-0.6}$ & -- & --\\
\hline
\end{tabular}
\end{center}
\label{tab:paramest}
\end{table}%

For fitting procedure, we assume Gaussian uncertainties since a full treatment of the experimental uncertainties is beyond the scope of this work. We then fit the event rate measurements by minimizing the $\chi^2$~
\be
\chi^2 = \sum_i{{\left(\tilde \sigma^i - {\tilde\Gamma_{\rm prod}^i \tilde\Gamma_{\rm decay}^i \over \tilde\Gamma_{\rm total}^i} \right)^2 \over (\delta \tilde \sigma^i)^2}},
\ee
where $\tilde \sigma$ and $\tilde \Gamma$ are the signal cross section and decay width scaled with respect to the SM expectation, respectively.  The measurement uncertainty on the cross section is given by $\delta \tilde \sigma$ and the asymmetric errors are retained.  

The outcome of the fits is summarized in Table~\ref{tab:paramest}, where we showed the $\chi^2$ per degree-of-freedom for a SM Higgs boson with all the effective couplings fixed at the SM value, a generic Higgs boson with free varying effective couplings, and a triplet imposter. We see that a generic Higgs boson gives the best fit among the three to the current data with a $p$-value of 0.27, while the SM Higgs and a triplet imposter give increasingly worse fits ($p$-values of 0.70 and 0.84, respectively).  We can also estimate the parameters of the Higgs boson and the triplet imposter at the $1\sigma$ level by determining the interval about which $\Delta \chi^2 \le 1$, also shown in Table~\ref{tab:paramest}. In the generic Higgs case, both $c_g$ and $c_V$ have best-fit values very close to the SM expectations, while $c_\gamma$ is significantly enhanced over the SM expectation. The best-fit $c_\tau=0$ is driven by the lack of excess in the CMS $\tau\tau$ measurement. In the triplet case, we generally find lower best-fit values of $c_g, c_V$ and $c_\gamma$.  This is expected as we assume the triplet does not decay into fermions and the corresponding total with is therefore smaller then in the Higgs case, which gives rise to larger branching fractions and lower production cross sections.

\begin{figure}[tpb]
\begin{center}
\includegraphics[scale=0.49, angle=0]{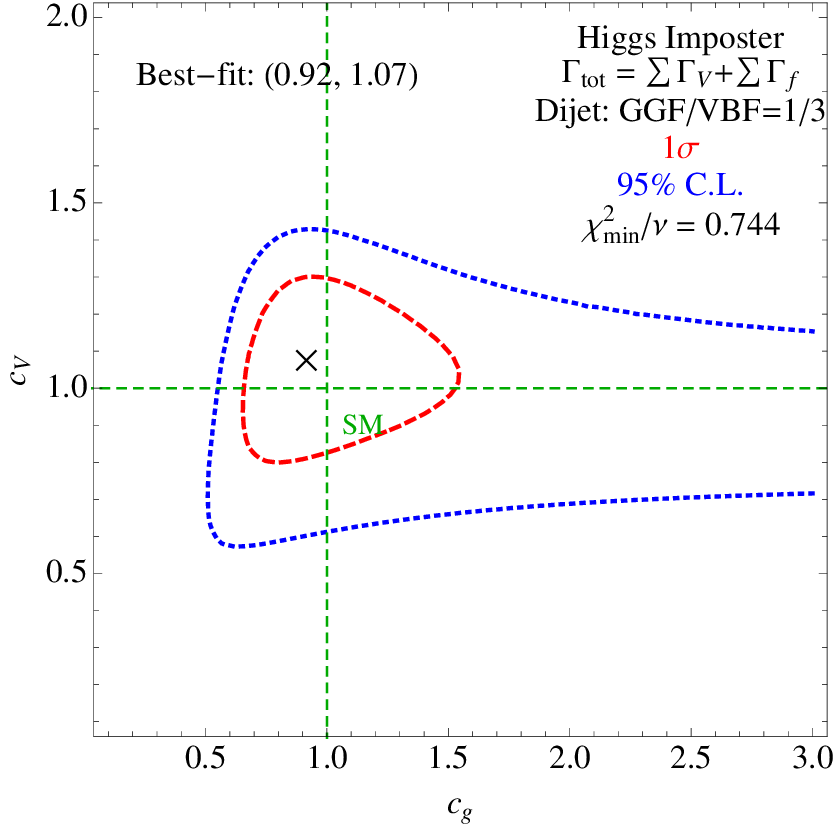}
\includegraphics[scale=0.49, angle=0]{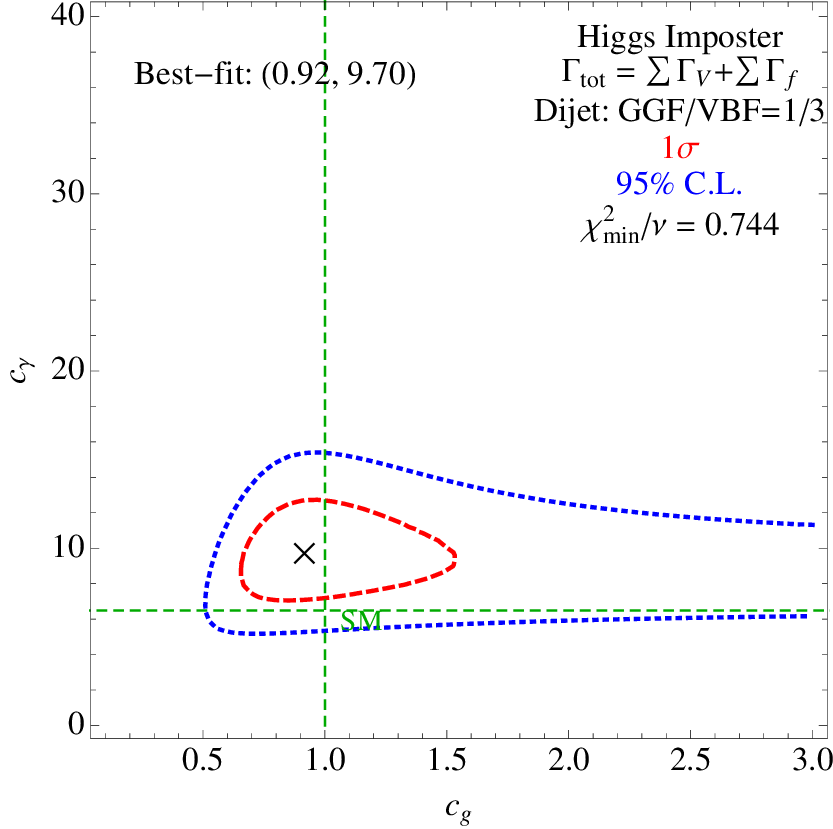}
\includegraphics[scale=0.49, angle=0]{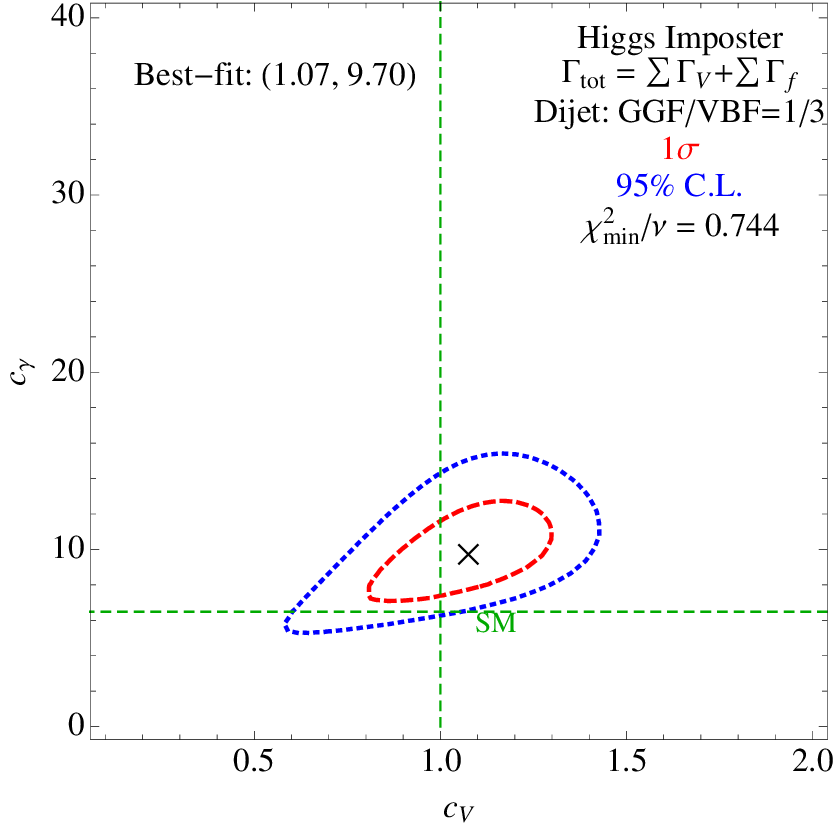}
\includegraphics[scale=0.49, angle=0]{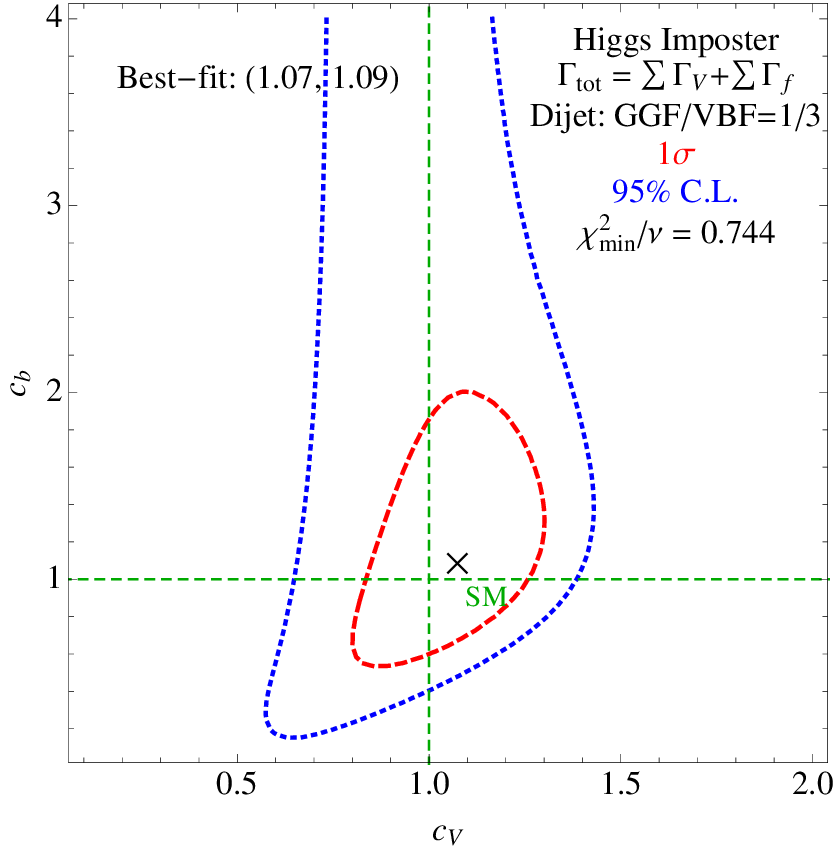}
\caption{\em Two-dimensional contours for four pairs of effective couplings. $(c_V, c_g)$ enters into the decays into $WW$ and $ZZ$ from gluon fusion production.
$(c_\gamma, c_g)$ enters into the decays into diphotons from gluon fusion production. $(c_\gamma, c_v)$ enters into the decays into diphotons from vector boson fusion production.
$(c_b, c_V)$ enters into the decays into $b\bar{b}$ from associated production with $W/Z$.}
\label{fig:2d_singlet}
\end{center}
\end{figure}

\begin{figure}[tpb]
\begin{center}
\includegraphics[scale=0.49, angle=0]{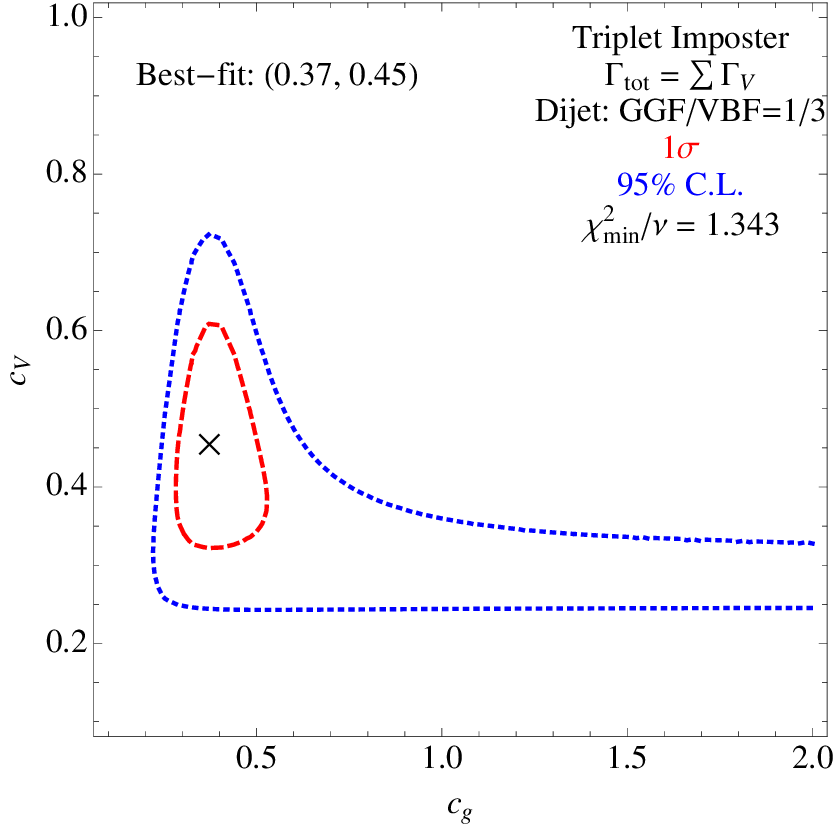}
\includegraphics[scale=0.49, angle=0]{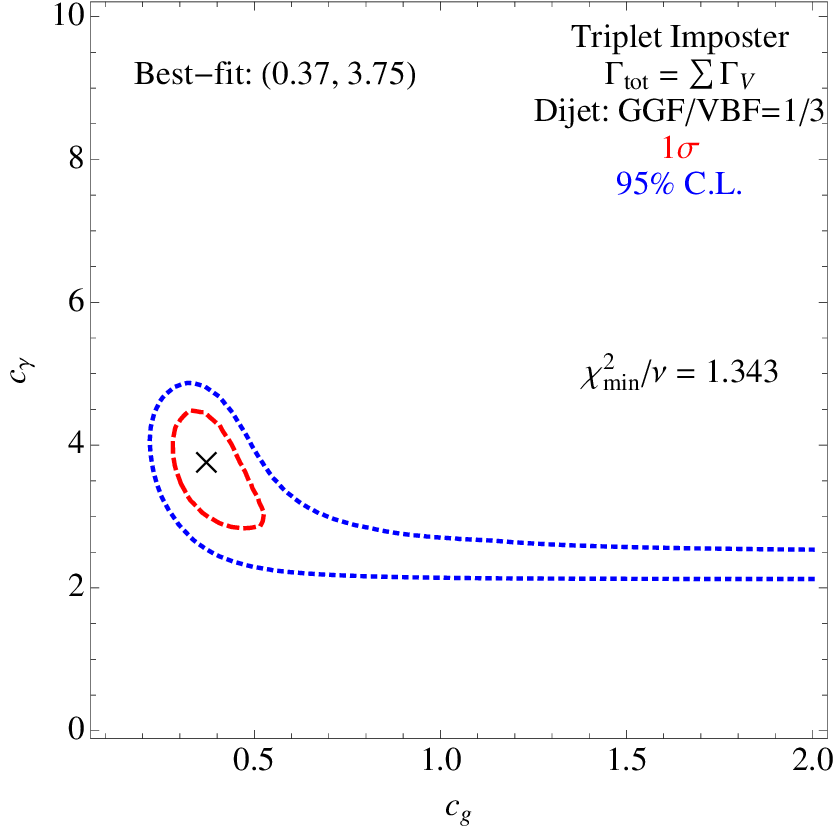}
\includegraphics[scale=0.49, angle=0]{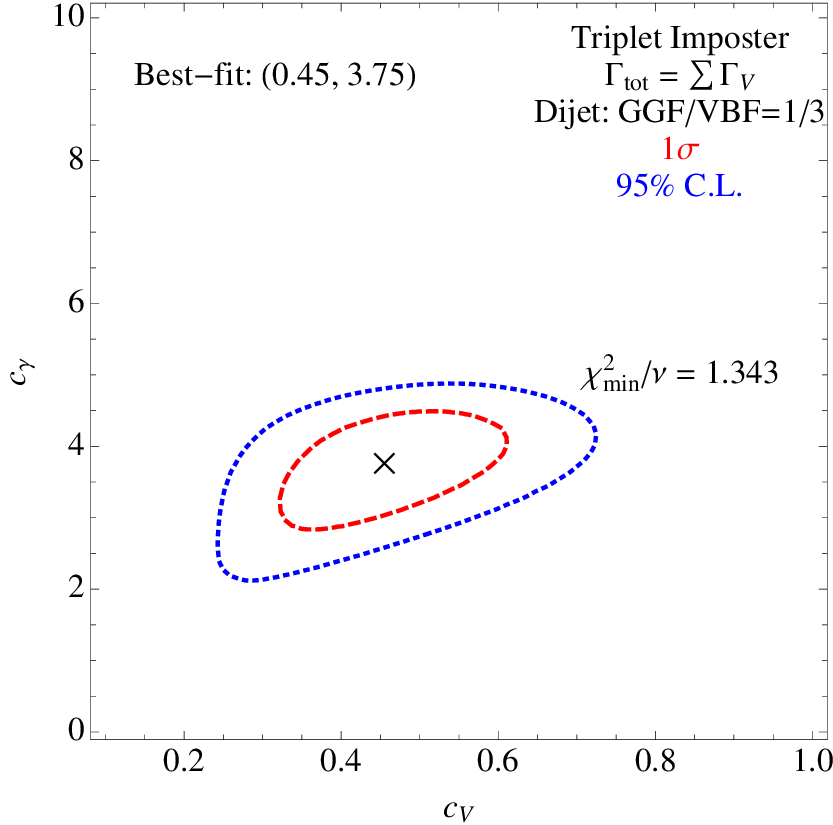}
\caption{\em Two-dimensional $\chi^2$ contours for the triplet imposter. There are only three pairs of effective couplings, which enter into the event rates in the $V_1V_2$ channel. The triplet imposter does not decay into $b\bar{b}$ and $\tau\tau$ final states by assumptions.}
\label{fig:2d_5plet}
\end{center}
\end{figure}
In Fig.~\ref{fig:2d_singlet}, we show the joint probability map in the plane of two model parameters.\footnote{Note, that since these are joint two-dimensional distributions, the $1\sigma$ region may lie outside the one parameter confidence intervals shown in Table~\ref{tab:paramest}.}  We choose to show the contours for the following four pairs of coefficients which enter into the channels with significant excesses: $(c_V, c_g)$ for $gg\to h \to WW/ZZ$, $(c_\gamma, c_g)$ for $gg\to h\to \gamma\gamma$, $(c_\gamma, c_V)$ for ${\rm VBF} \to h \to \gamma\gamma$, and $(c_b, c_V)$ for $Vh\to V+b\bar{b}$. Since production primarily occurs through $gg$ fusion and VBF, we expect to see the values of $c_g$ and $c_V$ to have a strong upper bound, while the value of $c_\gamma$ is allowed to rise well beyond the SM value of $c_\gamma^{\rm SM}=6.48$.  However, since  $gg$ fusion can contaminate the dijet channel, the value of $c_g$ can rise to compensate for a lower $c_V$.  Indeed, at the 2$\sigma$ level, the value of $c_g$ can be quite large.  We generally find good agreement with the SM expectation, with the exception to $c_\gamma$.  Indeed, the one dimensional parameter fit of $c_\gamma$ is nearly $2\sigma$ away from the SM value.  Overall, to fit the data, we require an enhancement to $\gamma\gamma$.

The corresponding two-dimensional contours for the triplet imposter are shown in Fig.~\ref{fig:2d_5plet}. In this scenario the resonance decays into vector bosons with no appreciable decay into $f\bar f$.  This is immediately at odds with the Tevatron $Vh\to V b\bar b$ result.  However, the absence of a signal in the CMS measurements in  ${\rm VBF}\to h\to \tau\tau$ channel supports this possibility.  Due to the absence of the fermonic decay modes, we expect the total width to be smaller than in the singlet case, which is consistent with the fits shown in Fig.~\ref{fig:2d_5plet}.  The value of $c_g$ is substantially lower than what is expected in the SM, meaning the total production of the scalar is suppressed.  However, the decay branching fraction to $\gamma\gamma$ and $WW/ZZ$ is increased due the lower total width.

\section{Conclusions}\label{sec:conclusions}

Under the assumption that the new resonance discovered at the LHC is a $CP$ even scalar particle with mass
125 GeV, we have performed a general analysis of its possible electroweak quantum numbers. We have
used a naive combination of the latest data from ATLAS, CMS, and the Tevatron experiments, focusing on
the four possible decays into pairs of electroweak gauge bosons, $\{WW, ZZ, \gamma\gamma, Z\gamma\}$,
but also taking into account the two most important decay channels into pairs of fermions,  $b\bar{b}$ and $\tau\tau$.

We have seen that interpreting the new particle as an electroweak singlet with loop-induced couplings to $W$ and $Z$ is strongly disfavored by current data. So is a "plain-vanilla" dilaton arising from scenarios where the SM gauge bosons are part of the conformal dynamics. It will be important for the
LHC experiments to quantify this statement, both by better constraints on decays to $Z\gamma$, and by
more accurate measurements of the VBF production modes. In the latter regard we note the critical importance
of having reliable estimates of the contamination of VBF analyses by $gg$ fusion-initiated signal events.

Using chi-squared fits to the relevant free parameters, we have compared the compatibility of current data
between a SM Higgs boson, a more general custodial singlet boson, and a custodial 5-plet boson as would arise
from an electroweak scalar triplet. All of the fits show some tension with the data, but the differences in the fit
quality are not large. Thus, for example, one can not yet exclude the possibility that the new particle is the
neutral member of electroweak triplets, provided that one is willing to discount the Tevatron excess in
$b\bar{b}$. Similarly one cannot greatly prefer a SM Higgs over a more general custodial singlet scalar,
especially if one takes seriously the lack of a $\tau\tau$ excess in the CMS data.
As we have seen, precise measurements of the ratios $D_{W/Z}$ and $D_{\gamma/Z}$ offer a clean
way of distinguishing a triplet imposter from a Higgs boson, but currently the uncertainties in these
quantities are too large, and the central values actually favor the triplet imposter.

\begin {acknowledgements}
We would like to thank Heather Logan, Alessandro Strumia, and Rik Yoshida for helpful correspondences.
This work was supported in part by the U.S. Department of Energy under
contracts No. DE-AC02-06CH11357 and No. DE-FG02-91ER40684. Fermilab is operated by the 
Fermi Research Alliance under contract DE-AC02-07CH11359 with
the U.S. Department of Energy.
\end{acknowledgements}

\end{document}